\documentclass[a4paper,12pt]{article}
\title{Nutations of spin in the quasi-isotropic superfluid A-like phase of $^3$He.}
\author {I.A.Fomin\\
P. L. Kapitza Institute for Physical Problems, \\
ul. Kosygina 2, 119334 Moscow,Russia}
\date{ }
\begin{document}
\maketitle
\begin{abstract}
The order parameter of the quasi-isotropic superfluid A-like phase
of $^3$He is rewritten in a simple form. Effect of anisotropy of
magnetic susceptibility of this phase on  the frequencies of the
spatially uniform oscillations of spin and the spin part of its
order parameter are found. Anisotropy of susceptibility has
pronounced effect on the low-frequency mode, which is analogous to
nutations of the asymmetric top. Possibility of observation of the
nutation-like mode is discussed.

\end{abstract}

\section {Introduction}
The order parameter of the superfluid A-like phase of $^3$He is
not yet identified. The measured magnetic susceptibility of the
A-like phase coincides with the susceptibility of the normal
phase. \cite{osher}. It means that the A-like phase belongs to the
ESP
 (Equal Spin Pairing) type and its order parameter can be written
 in a form:
$$
A^{ESP}_{\mu j}=\Delta\frac{1}{\sqrt{3}}[\hat d_{\mu}( m_j+i n_j)+
\hat e_{\mu}( l_j+i p_j)]. \eqno(1)
$$
 Here $\hat d_{\mu}$ and $\hat e_{\mu}$ are mutually orthogonal
 unit vectors, $m_j,n_j,l_j,p_j$ -- arbitrary real vectors
 normalized as:
 ${\bf m}^2+{\bf n}^2+{\bf l}^2+{\bf p}^2=3$.
A-like phase is not ferromagnetic. The minima of the free energy
of the bulk (free of impurities) $^3$He in a vicinity of $T_c$:
$$
f=f_n+\alpha A_{\mu j}A_{\mu j}^*+\beta_1 |A_{\mu j}A_{\mu
j}|^2+\beta_2(A_{\mu j}A_{\mu j}^*)^2+
$$
$$
\beta_3A_{\mu j}^*A_{\nu j}^*A_{\nu l}A_{\mu l}+ \beta_4A_{\mu
j}^*A_{\nu j}A_{\nu l}^*A_{\mu l}+ \beta_5A_{\mu j}^*A_{\nu
j}A_{\nu l}A_{\mu l}^*. \eqno(2)
$$
corresponding to nonferromagnetic ESP phases were found by Mermin
and Stare \cite{mermin}.
 There are four such minima (phases): axial (ABM),
 polar, bipolar and axiplanar. All four order parameters can be
 represented in a form:
$$
A^{0}_{\mu j}=\Delta e^{i\phi}[\hat d_{\mu}(v_x\hat m_j+iv_y\hat
n_j)+ v_z\hat e_{\mu}\hat l_j],  \eqno(3)
$$
where  $\hat d_{\mu}$ and $\hat e_{\mu}$ are mutually orthogonal
unit  vectors in spin space, $\hat{\bf l},\hat{\bf m},\hat{\bf n}$
- three orthonormal vectors in momentum space,  $v_x, v_y, v_z$ -
real numbers restricted by the condition: $v_x^2+v_y^2+v_z^2=1$.
Every phase is specified by a set of coefficients  $v_x, v_y,
v_z$. For the ABM phase these are $v^2_x=v^2_y=1/2, v_z=0$. In a
vicinity of $T_c$ effect of aerogel on the order parameter can be
taken into account phenomenologically by introduction in the
Ginzburg and Landau functional of additional term
 $\eta_{jl}({\bf r})A_{\mu j}A_{\mu l}^*$. A random
 symmetric tensor $\eta_{jl}({\bf r})$ is contracted with the orbital
 indices of the matrices $A_{\mu j}$ and $A_{\mu l}^*$. The order
 parameters of all phases are continuously degenerate with
 respect to orbital rotations. According to the general statement of
 Imry and Ma \cite{imry} in that case the interaction $\eta_{jl}({\bf r})$
 dirsrupts the orientational long-range order.
 Volovik \cite{volov1} considered the situation when original order
 parameter is the ABM. He arrived at the result that the disruption of
 the orientational long-range order leads to a glass type state with a
 short-range order described by the ABM order parameter. The
 resulting ordering is described by the nonvanishing averages
 $Q_{\mu\nu jl}=<A_{\mu j}({\bf r})A^*_{\nu l}({\bf
r})>$ and according to the Ref. \cite{volov1} corresponds to a
nonsuperfluid spin nematic.

 There is an additional effect of aerogel
on the superfluid $^3$He. The random field $\eta_{jl}({\bf r})$
induces local fluctuations of the order parameter, which in their
turn bring additional  contribution to the free energy. As a
result the total free energy can have minima, which are different
from the listed above. When the contribution of fluctuations is
dominating e.g. in a vicinity of $T_c$, the quasi isotropic (or
"robust") A-like phase with a long-range, or "nearly long-range"
order becomes a minimum \cite{fom1}. This phase remains superfluid
even when orientational long-range order is disrupted and a
glass-like state is formed. The choice between different
possibilities depends on concrete parameters of the system, in
particular on values of the coefficients $\beta_1,...\beta_5$ in
Eq. (2). Because of a lack of precise knowledge of these
coefficients even in the bulk $^3$He the choice between different
possibilities existing in aerogel can not be made on a basis of
purely theoretical argument. In such situation for the
identification of this phase one has to rely on the comparison of
the observed properties of the A-like phase with the predicted
theoretically can be of a great importance .

Aerogel effects directly the orbital part of the order parameter
of $^3$He.The spin part of the order parameter is weakly coupled
with the orbital via the dipole interaction. On the other hand the
spin part interacts with magnetic field and directly manifests
itself in NMR experiments. These properties make NMR a convenient
"probe" of the superfluid state and of its order parameter.

Among the existing NMR data only one experiment \cite{ishik} makes
direct distinction between the ABM and quasi isotropic phase. In
that pulsed NMR experiment the dependence of the transverse
frequency shift on the tipping angle has been measured. The
measured dependence coincides with that predicted for the "robust"
phase \cite{miura} and does not coincide with the expected for the
ABM phase. The bulk of data of the NMR experiments with the A-like
phase in not yet completely understood. In particular the origin
of the observed in many experiments negative cw-NMR  shift is not
clear. In this situation the agreement of one experiment with the
theoretical prediction can not be taken as a final proof of the
suggested identification. Further experiments would be very useful
in making the identification definitive.

A possible experiment could be observation of the low-frequency
transverse  NMR mode which according to Ref.\cite{miura} has to be
present in the quasi-isotropic phase but is absent in the ABM
phase. In the Ref.\cite{miura} the frequencies of the NMR modes
were found without account of anisotropy of magnetic
susceptibility of the quasi-isotropic phase. It will be shown
later that this anisotropy changes substantially the frequency of
the above mentioned  mode. In the present paper frequencies of all
NMR modes for the quasi-isotropic phase are found with account of
the anisotropy of susceptibility and a possibility of observation
of the low frequency mode is discussed. In comparison with the
Ref.\cite{miura} calculations are simplified because the order
parameter of the quasi-isotropic phase is represented in a more
convenient form.

\section {The order parameter}
Quasi-isotropic phases are determined by the condition
\cite{fom1}:
$$
\eta^{(a)}_{jl}A_{\mu j}A_{\mu l}^*=0, \eqno(4)
$$
where $\eta^{(a)}_{jl}$ is an arbitrary real symmetric traceless
tensor. Eq.(4) is equivalent to:
$$
 A_{\mu l} A^*_{\mu j}+A_{\mu j} A^*_{\mu l}=
const\cdot\delta_{lj}.  \eqno(5)
$$
The order parameter (1) meets criterion (5) if  $m_j,n_j,l_j,p_j$
satisfy the following equation:
$$
m_jm_l+n_jn_l+l_jl_l+p_jp_l=\delta_{jl}. \eqno(6)
$$
The order parameter is degenerate with respect to separate
rotations in spin and momentum spaces. For that reason any member
of this degenerate family can be chosen as a representative. It
has been shown before that the vectors
 ${\bf m},{\bf n},{\bf l},{\bf p}$ satisfying Eq.(6) have the following property:
 $[{\bf m}\times{\bf n}]\cdot[{\bf l}\times{\bf p}]=0$. That makes possible to
 introduce three mutually orthogonal unit vectors
 $\hat{\bf a},\hat{\bf b},\hat{\bf c}$ in the following way:
$\hat{\bf a}\|({\bf m}\times{\bf n}),\hat{\bf b}\|({\bf
l}\times{\bf p}),\hat{\bf c}=\hat{\bf a}\times\hat{\bf b}$. Only
non ferromagnetic phases will be considered. Expressing
 ${\bf m},{\bf n},{\bf l},{\bf p}$ in terms of
 $\hat{\bf a},\hat{\bf b},\hat{\bf c}$ and after straightforward
 transformations one arrives at:
$$
A^R_{\mu j}=\Delta\frac{1}{\sqrt{3}}e^{i\psi}[\hat d_{\mu}(\hat
b_j+i\cos\gamma\hat c_j)+ \hat e_{\mu}(\hat a_j+i\sin\gamma\hat
c_j)], \eqno(7)
$$
where $\psi$ and $\gamma$ are arbitrary angles. Let us express
vectors $\hat{\bf d},\hat{\bf e}$ in terms of the new vectors
$\hat{\bf d'},\hat{\bf e'}$ obtained by rotation of the pair
 $\hat{\bf d},\hat{\bf e}$ around
$\hat{\bf f}=\hat{\bf d}\times\hat{\bf e}$ for an angle  $\gamma$,
and vectors $\hat{\bf a},\hat{\bf b},\hat{\bf c}$ in terms of
$\hat{\bf l},\hat{\bf m},\hat{\bf n}$ obtained by rotation of
$\hat{\bf a},\hat{\bf b}$ around $\hat{\bf c}$ for $-\gamma$. The
hats are introduced to distinguish the unit vectors  $\hat{\bf
l},\hat{\bf m},\hat{\bf n}$ from the vectors $l_j,m_j,n_j$,
entering eqns. (3),(4).After the above transformations the order
parameter assumes the simple form:
$$
A^R_{\mu j}=\Delta\frac{1}{\sqrt{3}}e^{i\psi}[\hat d_{\mu}(\hat
m_j+i\hat n_j)+ \hat e_{\mu}\hat l_j].  \eqno(8)
$$
(The primes at $\hat{\bf d'},\hat{\bf e'}$ are omitted).  $A_{\mu
j}^R$ can also be represented in a form (2) with the coefficients
 $v_x^2=v_y^2=v_z^2=1/3$. Each of the coefficients
 $v_x, v_y, v_z$ assumes two values $\pm\frac{1}{\sqrt{3}}$.
Without magnetic field all combinations of signs yield equivalent
order parameters, i.e. transformed into each other by a
combination of rotations in spin and orbital spaces. The
particular combination of signs, chosen in Eq. (8) in that sense
is representative.

\section {Nutation mode}
It is shown in Ref.\cite{miura} that in magnetic field
qausi-isotropic phase has three modes of small oscillations at the
equilibrium configuration: one longitudinal and two transverse. If
the Larmor frequency $\omega_L$  is much greater then the dipole
frequency $\Omega$, and if anisotropy of the magnetic
susceptibility is not taken into account, the frequency of one of
the transverse modes is close to the Larmor frequency and is of
the other $\Omega^2/\omega_L$.

For the order parameter Eq.(8)  in a vicinity of $T_c$ the
magnetic susceptibility tensor has a form:
$$
\chi_{\mu \nu}=\chi_n[\delta_{\mu
\nu}-\epsilon(2d_{\mu}d_{\nu}+e_{\mu}e_{\nu})],    \eqno(9)
$$
where $\epsilon$ is a positive coefficient $\epsilon\sim(T-
T_c)/T_c$. The maximum eigenvalue $\chi_n$ corresponds to the
direction $\hat{\bf f}=\hat{\bf d}\times\hat{\bf e}$, i.e. in the
equilibrium  $\hat{\bf f}$ is parallel or antiparallel to the
magnetic field.

The equilibrium configuration of the order parameter in magnetic
field and with the account of the dipole interaction was
determined before in terms of the vectors  $m_j,n_j,l_j,p_j$
\cite{miura,fom3}. Here this configuration is reformulated for the
order parameter in the form Eq. (8). Orientation of $\hat{\bf
l},\hat{\bf m},\hat{\bf n}$ with respect to $\hat{\bf d},\hat{\bf
e},\hat{\bf f}$ is determined by minimization of the dipole
energy:
$$
U_D=\frac{\chi_n}{8 g^2}\Omega^2\left[(\hat{\bf d}\cdot\hat{\bf
m}+\hat{\bf e}\cdot\hat{\bf l})^2+(\hat{\bf d}\cdot\hat{\bf
n})^2+\hat{\bf f}\cdot\hat{\bf n}\right]. \eqno(10)
$$
Coefficient in front of the energy is expressed in terms of the
longitudinal oscillations frequency $\Omega$ and of the
gyromagnetic ratio  $g$. Minimum of $U_D$ is reached at $\hat{\bf
d}\cdot\hat{\bf m}+\hat{\bf e}\cdot\hat{\bf l}=0$, $\hat{\bf
d}\cdot\hat{\bf n}=0$, $\hat{\bf f}\cdot\hat{\bf n}=-1$. According
to the last condition  $\hat{\bf f}$ and $\hat{\bf n}$ are
antiparallel in the equilibrium. Two other conditions determine
two possible equilibrium orientations of   $\hat{\bf l},\hat{\bf
m}$ with respect to  $\hat{\bf d},\hat{\bf e}$ in a plane
perpendicular to
 $\hat{\bf f}$. These are $\hat{\bf
l}=\hat{\bf d}$, $\hat{\bf m}=-\hat{\bf e}$ and $\hat{\bf
l}=-\hat{\bf d}$, $\hat{\bf m}=\hat{\bf e}$. The dipole
interaction lifts the degeneracy with respect to arbitrary
rotations of the orbital vectors  $\hat{\bf l},\hat{\bf
m},\hat{\bf n}$ relative to the spin vectors $\hat{\bf d},\hat{\bf
e},\hat{\bf f}$. Only a discreet two-fold degeneracy is preserved.
The discreet degeneracy is analogous to that in the ABM-phase
 $\hat{\bf l}=\pm\hat{\bf d}$.

Frequencies of cw-NMR are found from the Leggett equations. With
the explicit form of the dipole energy (10) and of the
susceptibility tensor  (9) these equations read as:
$$
\dot{\bf S}={\bf S}\times\vec{\omega}_L-\frac{\chi_n\Omega^2}{8
g^2} \{2(\hat{\bf d}\cdot\hat{\bf m}+\hat{\bf e}\cdot\hat{\bf l})
(\hat{\bf d}\times\hat{\bf m}+\hat{\bf e}\times\hat{\bf l})+
2(\hat{\bf d}\cdot\hat{\bf n})\hat{\bf d}\times\hat{\bf n}+
\hat{\bf f}\times\hat{\bf n}\}  \eqno (11)
$$

$$
\dot{\vec{\theta}}=\frac{g^2}{\chi_n}[{\bf S}+\zeta_1\hat{\bf d}
(\hat{\bf d}\cdot{\bf S})+\zeta_2\hat{\bf e} (\hat{\bf e}\cdot{\bf
S})]-\vec{\omega}_L. \eqno (12)
$$
Here  $\vec{\omega}_L=g{\bf H}$ and $\dot{\vec{\theta}}$-- the
angular velocity of the spin vectors  $\hat{\bf d},\hat{\bf
e},\hat{\bf f}$ i.e. $\hat{\dot{\bf
d}}=\dot{\vec{\theta}}\times\hat{\bf d}$ etc.. Tensor of inverse
susceptibilities is written as:
$$
\chi^{-1}_{\mu \nu}=\frac{1}{\chi_n}[\delta_{\mu \nu}+\zeta_1
d_{\mu}d_{\nu}+\zeta_2 e_{\mu}e_{\nu}],    \eqno(13)
$$
where $\zeta_1=2\epsilon/(1-2\epsilon)$ and
$\zeta_2=\epsilon/(1-\epsilon)$. Linearization of Eqns.(11),(12)
at the equilibrium  $\hat{\bf f}\parallel{\bf H}$, $\hat{\bf
n}=-\hat{\bf f}$, $\hat{\bf l}=\hat{\bf d}$, $\hat{\bf
m}=-\hat{\bf e}$ renders equations for three modes of harmonic
oscillations --  longitudinal with the frequency
$$
\omega^2_{\parallel}=\Omega^2 \eqno(14)
$$
and two transverse with the frequencies $\omega_{\pm}$ determined
by the equation:
$$
2\omega^2_{\pm}=\omega_L^2(1+\zeta_1\zeta_2)+\frac{\Omega^2}{8}
(4+\zeta_1+3\zeta_2)\pm
$$
$$
\sqrt{[\omega_L^2(1-\zeta_1\zeta_2)+ \frac{\Omega^2}{8}
(4-\zeta_1-3\zeta_2)]^2+\frac{\Omega^4}{16}(\zeta_1-3)(1-3\zeta_2)}.
\eqno(15)
$$
In the absence of anisotropy of the susceptibility ($\epsilon=0$),
the frequencies (14), (15) coincide with that found before
\cite{miura}.

NMR experiments with the A-like phase are usually performed in
magnetic fields for which the condition $\omega_L\gg\Omega$ is
met. Expansion over the small ratio $\Omega/\omega_L$ renders less
cumbersome expressions for the frequencies of the transverse
modes. In zero order on  $\Omega/\omega_L$ the frequency
$\omega^2_+=\omega_L^2$, it corresponds to the Larmor precession
of spin. The other frequency
$$
\omega^2_-=\omega_L^2\zeta_1\zeta_2=\omega_L^2\frac{\chi_{ff}-\chi_{dd}}
{\chi_{dd}}\frac{\chi_{ff}-\chi_{ee}} {\chi_{ee}},  \eqno(16)
$$
where  $\chi_{ff}=\chi_n$ is the maximum,  $\chi_{dd}$ and
$\chi_{ee}$ -- two other principal values of the tensor of
magnetic susceptibility, corresponds to the motion of the spin
vectors $\hat{\bf d},\hat{\bf e},\hat{\bf f}$, which is analogous
to \emph{nutations}  of a classical asymmetric top \cite{landau}.
Taking into consideration further terms of the expansion on
$\Omega/\omega_L$ one obtains for the square of the Larmor-like
mode:
$$
\omega^2_+=\omega_L^2+\frac{\Omega^2}{2}-\frac{\Omega^4}{64\omega_L^2}
\frac{(3-\zeta_1)(1-3\zeta_2)}{1-\zeta_1\zeta_2}, \eqno(17)
$$
i.e.  corrections  due to the anisotropy of susceptibility start
from the terms of the relative order  $(\Omega/\omega_L)^4$. For
nutation-like mode correction of the order of $\Omega^2$ is:
$$
\omega^2_-=\omega_L^2\zeta_1\zeta_2+\frac{\Omega^2}{8}(\zeta_1+3\zeta_2)+
O\left(\frac{\Omega^4}{\omega_L^2}\right),  \eqno(18)
$$
The first two terms in the r.h.s of Eq. (18) disappear when
anisotropy tends to zero. The first term disappears also when the
maximum principal value of the tensor of magnetic susceptibility
coincides with one of the the remaining principal values.

For the pulsed NMR the frequency shift as a function of the
tipping angle in a principal order on $\Omega/\omega_L$  can be
found with the aid of a standard procedure  \cite{fom4}.
Instantaneous orientation of vectors  $\hat{\bf d},\hat{\bf
e},\hat{\bf f}$ is determined by the Euler angles
$\alpha,\beta,\gamma$ according to the definition: $\hat{\bf
d}(t)=R_z(\alpha)R_y(\beta)R_z(\gamma)\hat{\bf d}_0$ etc., where
$\hat{\bf d}_0$  is an equilibrium orientation of  $\hat{\bf d}$,
and the axes  $x,y,z$ are directed along  $\hat{\bf d}_0,\hat{\bf
e}_0,\hat{\bf f}_0$ correspondingly. The dipole energy (10) has to
be expressed in terms of  $\alpha,\beta,\gamma$
$$
U_D=\frac{\chi_n}{8
g^2}\Omega^2\left[(1+\cos\beta)^2\sin^2(\alpha+\gamma)+\sin^2\beta\cos^2\gamma-
\cos\beta\right]. \eqno(19)
$$
and to be averaged over the "fast" variables  $\alpha$ and
$\gamma$ with the fixed combination $\phi=\alpha+\gamma$. That
renders:
$$
V=\bar U_D=\frac{\chi_n}{8
g^2}\Omega^2\left[(1+\cos\beta)^2\sin^2\phi+\frac{1}{2}\sin^2\beta-
\cos\beta\right]. \eqno(20)
$$
The pulsed NMR shift $\omega_{\bot}(\beta)-\omega_L$ is determined
by the derivative of $(-V/\chi_nH_0^2)$ over $\cos\beta$  in the
minimum of  $V$ over $\phi$. The obtained dependence:
$$
\omega_{\bot}(\beta)=\omega_L+\frac{\Omega^2}{8\omega_L}(1+\cos\beta)
\eqno(21)
$$
coincides with the obtained before \cite{miura}.

  Unfortunately the observation of
the mode (18) is not an easy task. In NMR experiments a motion of
magnetization  (or spin) is registered. If the dipole energy is
neglected spin according to Eqns. (11),(12) precesses with the
Larmor frequency irrespective of whether  nutations are excited or
not. Transverse magnetic field does not interact with the
nutations. The coupling is provided only by the dipole
interaction. As a result, the intensity of the corresponding line
in the NMR spectrum has to be proportional to a power of the ratio
$\Omega/\omega_L$. For a quantitative evaluation of this intensity
one can apply the expression for absorption of the power $P$ from
the r.f. field
 ${\bf H}_1(t)={\bf H}_1\cos(\omega t)$:
 $$
 P=\frac{1}{4}i\omega
 H_{1\alpha}H_{1\beta}[\chi_{\alpha\beta}(-\omega)-\chi_{\alpha\beta}(\omega)].
 \eqno(22)
 $$
The answer depends on the polarization of ${\bf H}_1(t)$ i.e. on
the orientation of the amplitude ${\bf H}_1$ with respect to the
orbital vectors  $\hat{\bf l},\hat{\bf m}$. For a random
orientation of $\hat{\bf l},\hat{\bf m}$ in the plane,
perpendicular to  ${\bf H}_0$  the trace
$\chi_{\alpha\alpha}(\omega)$ enters the expression for intensity.
The trace has poles at $\omega=\pm\omega_+$  and
$\omega=\pm\omega_-$. The ratio of intensities $I$  of the two
lines is determined by the ratio of  residues in the corresponding
poles:
$$
\frac{I_-}{I_+}=\frac{\omega_-Res\chi_{\alpha\alpha}(\omega_-)}
{\omega_+Res\chi_{\alpha\alpha}(\omega_+)}.   \eqno(23)
$$
To find the residues one has to substitute  ${\bf H}_1(t)$ in
Eqns. (11), (12) and to find a linear response to this field.
After straightforward calculations one arrives at the expression:
$$
\frac{I_-}{I_+}=\frac{(1+2\psi)[1-\zeta_1\zeta_2+\psi(4-\zeta_1-3\zeta_2)-R]
-\psi^2(6-\zeta_1-9\zeta_2)}
{(1+2\psi)[1-\zeta_1\zeta_2+\psi(4-\zeta_1-3\zeta_2)+R]
-\psi^2(6-\zeta_1-9\zeta_2)}. \eqno(24)
$$
Shorthand notations:$\psi=\Omega^2/8\omega_L^2$ and
$R=\{[1-\zeta_1\zeta_2+\psi(4-\zeta_1-3\zeta_2)]^2
-4\psi^2(3-\zeta_1)(1-3\zeta_2)\}^{1/2}$ are introduced here. The
leading term of the ratio of intensities at  $\psi\ll 1$ is:
$$
\frac{I_-}{I_+}=\psi^2\frac{\zeta_1+3\zeta_2-12\zeta_1\zeta_2+\zeta_1^2\zeta_2
+9\zeta_1\zeta_2^2}{2(1-\zeta_1\zeta_2)^2}. \eqno(25)
$$
With account of  $\zeta_1, \zeta_2\ll 1$, a simple expression is
obtained:
$$
\frac{I_-}{I_+}=\frac{\Omega^4}{128\omega_L^4}(\zeta_1+3\zeta_2).
 \eqno(26)
$$
It shows, that the relative intensity of the nutation line is
going down with the increase of  $\omega_L$. Observation of this
mode may become possible in relatively weak fields
$\omega_L\sim\Omega$.
\section{Conclusions}
The low frequency "nutation" mode is a qualitative distinction of
the quasi-isotropic phase from the ABM-phase. In view of the above
estimation the absence of the corresponding line in experiments
can not be considered as an evidence against the quasi-isotropic
phase. In the experiments the condition $\omega_L\gg\Omega$ has
been well satisfied. In that sense NMR experiments with the
superfluid $^3$He in aerogel in fields $\omega_L\sim\Omega$ would
be useful.

In many experiments with the A-like phase a negative shift of the
NMR resonance frequency has been observed. Existence of absorption
at frequencies below the Larmor frequency indicates that
orientation of the spin triad  $\hat{\bf d},\hat{\bf e},\hat{\bf
f}$ relative to the orbital $\hat{\bf l},\hat{\bf m},\hat{\bf n}$
does not correspond to a minimum of the dipole energy. Deviation
from the minimum can occur because of  a presence of singularities
of the order parameter, in particular of domain walls. Possible
structures of the domain walls depend on the order parameter. In
the ABM-phase there are walls across which the relative
orientation of the vectors  $\hat{\bf d}$  and $\hat{\bf l}$
changes from $\hat{\bf d}\parallel\hat{\bf l}$   to $\hat{\bf
d}\parallel(-\hat{\bf l})$. In the quasi-isotropic phase there are
analogous domain walls with a continuous transition between two
minima of the dipole energy:  $\hat{\bf l}=\hat{\bf d}$, $\hat{\bf
m}=-\hat{\bf e}$ and $\hat{\bf l}=-\hat{\bf d}$, $\hat{\bf
m}=\hat{\bf e}$.

In a presence of magnetic field the quasi-isotropic phase has
additional type of the domain wall. The maximum principal value of
the magnetic susceptibility corresponds to the direction $\hat{\bf
f}$. In magnetic field ${\bf H}$ it is possible to have domains
with the different orientation of $\hat{\bf f}$: $\hat{\bf
f}\parallel{\bf H}$ and $\hat{\bf f}\parallel(-{\bf H})$,
separated by the domain wall with a thickness $\xi_H\sim
c/\omega_L$, where $c$ is the spin wave velocity. Minimum of the
dipole energy is reached at $\hat{\bf n}\parallel -{\bf f}$. The
orbital triad  $\hat{\bf l},\hat{\bf m},\hat{\bf n}$  is adjusted
to the spin in a layer of the thickness $\xi_D\sim c/\Omega$.
Within this layer mutual orientation of the spin and the orbital
vectors does not correspond to a minimum of the dipole energy.

Both in the ABM and in the quasi-isotropic phase there are reasons
for the negative shift of the NMR-line. Different defects manifest
themselves in a different shapes of the NMR-line, but
interpretation of the influence of defects requires rather
involved analysis and is not direct. The resulting conclusions
based on such interpretation are not as convincing as that based
on interpretation of NMR spectra of the uniform liquid.

Useful discussions with Dmitriev are gratefully acknowledged. This
work is partly supported by RFBR (grant 04-02-16417), Ministry of
Science and Education of the Russian Federation and CRDF (grant
RUP1-2632-MO04).

\end{document}